\documentclass[journal]{IEEEtran}

\usepackage{amsmath} 
\usepackage{amssymb}
\usepackage{subfigure}
\usepackage{color}
\usepackage{hyperref}
\usepackage{graphicx}

\usepackage{hyperref}
\usepackage{glossaries}
\usepackage{comment}
\usepackage{tabularx}

\newacronym{sc-rnti}{SC-RNTI}{Single-Cell Radio Network Temporary Identifier}
\newacronym{p-rnti}{P-RNTI}{Paging Radio Network Temporary Identifier}
\newacronym{uicc}{UICC}{Universal Integrated Circuit Card}
\newacronym{euicc}{eUICC}{embedded Universal Integrated Circuit Card}
\newacronym{musim}{MUSIM}{Multiple USIMs}
\newacronym{ims}{IMS}{IP Multimedia Subsystem}
\newacronym{cm}{CM}{Connection Management}
\newacronym{nas}{NAS}{Non-access Stratum}
\newacronym{amf}{AMF}{Access and Mobility Management Function}
\newacronym{3gpp}{3GPP}{Third Generation Partnership Project}
\newacronym{lte}{LTE}{Long Term Evolution}
\newacronym{ue}{UE}{User Equipment}
\newacronym{4g}{4G}{Fourth Generation}
\newacronym{5g}{5G}{Fifth Generation}
\newacronym{sim}{SIM}{Subscriber Identification Modules}
\newacronym{msma}{MSMA}{Multiple SIM Multiple Active}
\newacronym{msms}{MSMS}{Multiple SIM Multiple Standby}
\newacronym{dsda}{DSDA}{Dual SIM Dual Active}
\newacronym{dsds}{DSDS}{Dual SIM Dual Standby}
\newacronym{tau}{TAU}{Tracking Area Update}
\newacronym{usim}{USIM}{Universal Subscriber Identity Module}
\newacronym{imei}{IMEI}{International Mobile Equipment Identifier}
\newacronym{imsi}{IMSI}{International Mobile Subscriber Identifier}
\newacronym{tmsi}{TMSI}{Temporal Mobile Subscriber Identifier}
\newacronym{id}{ID}{Identity}
\newacronym{eps}{EPS}{Evolved Packet System}
\newacronym{5gs}{5GS}{5G System}
\newacronym{plmn}{PLMN}{Public Land Mobile Network}
\newacronym{sms}{SMS}{Short Message Service}
\newacronym{mno}{MNO}{Mobile Network Operator}
\newacronym{enb}{eNB}{evolved Node B}
\newacronym{ng-ran}{NG-RAN}{Next Generation RAN}
\newacronym{rrc}{RRC}{Radio Resource Control}
\newacronym{mm}{MM}{Mobility Management}
\newacronym{bs}{BS}{Base Station}
\newacronym{rat}{RAT}{Radio Access Technology}
\newacronym{sm}{SM}{Session Management}
\newacronym{rm}{RM}{Registration Management}
\newacronym{emm}{EMM}{EPS Mobility Management}
\newacronym{ecm}{ECM}{EPS Connection Management}
\newacronym{epc}{EPC}{Evolved Packet Core}
\newacronym{mme}{MME}{Mobility Management Entity}
\newacronym{e-utran}{E-UTRAN}{Evolved Universal Terrestrial Radio Access Network}
\newacronym{ran}{RAN}{Radio Access Network}
\newacronym{5gc}{5GC}{5G Core}
\newacronym{ngap}{NGAP}{Next Generation Application Protocol}
\newacronym{nr}{NR}{New Radio}
\newacronym{tai}{TAI}{Tracking Area Information}
\newacronym{pci}{PCI}{Physical Cell Identity}
\newacronym{cn}{CN}{Core Network}
\newacronym{po}{PO}{Paging Opportunity}
\newacronym{pf}{PF}{Paging Frame}
\newacronym{sgw}{SGW}{Serving Gateway}
\newacronym{gnb-du}{gNB-DU}{gNB Distributed Unit}
\newacronym{gnb-cu}{gNB-CU}{gNB Control Unit}
\newacronym{gsm}{GSM}{Global System for Mobile}
\newacronym{3g}{3G}{Third Generation}
\newacronym{umts}{UMTS}{Universal Mobile Telecommunications System}
\newacronym{hspa}{HSPA}{High Speed Packet Access}
\newacronym{cdma}{CDMA}{Code Division Multiple Access}
\newacronym{csim}{CSIM}{CDMA SIM}
\newacronym{isim}{ISIM}{IP Multimedia SIM}
\newacronym{ip}{IP}{Internet Protocol}
\newacronym{2g}{2G}{Second Generation}
\newacronym{mt}{MT}{Mobile Terminated}
\newacronym{mo}{MO}{Mobile Originated}
\newacronym{pdu}{PDU}{Protocol Data Unit}
\newacronym{supi}{SUPI}{5G Subscription Permanent Identifier}
\newacronym{5g-guti}{5G-GUTI}{5G Globally Unique Temporary Identifier}
\newacronym{guti}{GUTI}{Globally Unique Temporary Identifier}
\newacronym{nf}{NF}{Network Function}
\newacronym{smf}{SMF}{Session Management Function}
\newacronym{suci}{SUCI}{Subscription Concealed Identifier}
\newacronym{5g-s-tmsi}{5G-S-TMSI}{5G S-Temporary Mobile Subscription Identifier}
\newacronym{as}{AS}{Access Stratum}
\newacronym{ta}{TA}{Tracking Area}
\newacronym{upf}{UPF}{User Plane Function}
\newacronym{ppi}{PPI}{Paging Policy Indicator}
\newacronym{dscp}{DSCP}{Differentiated Services Code Point}
\newacronym{tos}{TOS}{Type of Service}
\newacronym{qos}{QoS}{Quality of Service}
\newacronym{qfi}{QFI}{QoS Flow Identifier}
\newacronym{arp}{ARP}{Allocation and Retention Priority}
\newacronym{5qi}{5QI}{5G QoS Identifier}
\newacronym{rau}{RAU}{Registration Area Update}
\newacronym{rnau}{RNAU}{Radio Network Area Update}
\newacronym{gnb}{gNB}{5G NodeB}
\newacronym{sgw-c}{SGW-C}{Serving Gateway for Control Plane}
\newacronym{sgw-u}{SGW-U}{Serving Gateway for User Plane}
\newacronym{cpu}{CPU}{Central Processing Unit}
\newacronym{drx}{DRX}{Discontinuous Reception}
\newacronym{s-tmsi}{S-TMSI}{System-Temporal Mobile System Identifier}
\newacronym{fqdn}{FQDN}{Fully Qualified Domain Name}
\newacronym{pws}{PWS}{Public Warning System}
\newacronym{sla}{SLA}{Service Level Agreement}
\newacronym{p-gw}{P-GW}{Packet Data Network Gateway}
\newacronym{s-gw}{S-GW}{Serving Gateway}
\newacronym{uu}{Uu}{User}
\newacronym{hss}{HSS}{Home Subscriber Server}
\newacronym{epdg}{ePDG}{Evolved Packet Data Gateway}
\newacronym{pdn}{PDN}{Public Data Network}
\newacronym{ausf}{AUSF}{Authentication Server Function}
\newacronym{pcf}{PCF}{Policy Control Function}
\newacronym{raa}{RA}{Registration Area}
\newacronym{dci}{DCI}{Downlink Control Information}
\newacronym{rna}{RNA}{Registration Network Area}
\newacronym{i-rnti}{I-RNTI}{Inactivity Radio Network Temporal Identifier}
\newacronym{andsf}{ANDSF}{Access Network Discovery and Selection Function}
\newacronym{esm}{ESM}{EPS Session Management}
\newacronym{msisdn}{MSISDN}{Mobile Subscriber Integrated Services Digital Number}
\newacronym{sms-sc}{SMS-SC}{SMS Service Center}
\newacronym{nef}{NEF}{Network Exposure Function}
\newacronym{scef}{SCEF}{Service Capability Exposure Function}
\newacronym{n3iwf}{N3IWF}{Non-3GPP Interworking Function}
\newacronym{esim}{eSIM}{embedded SIM}
\newacronym{gsma}{GSMA}{Global System for Mobile Association}

\definecolor{amethyst}{rgb}{0.6, 0.4, 0.8}
\definecolor{cobalt}{rgb}{0.0, 0.28, 0.67}
\definecolor{fashionfuchsia}{rgb}{0.96, 0.0, 0.63}
\definecolor{electricviolet}{rgb}{0.56, 0.0, 1.0}
\definecolor{grey}{rgb}{0.52, 0.52, 0.51}

\makeatletter
\def\ps@IEEEtitlepagestyle{%
	\def\@oddfoot{\mycopyrightnotice}%
	\def\@evenfoot{}%
}

\makeatother
\PassOptionsToPackage{draft}{hyperref}
\hypersetup{draft}

\usepackage{tikz}
    \usetikzlibrary{shapes.arrows}
    \usetikzlibrary{plotmarks}
    \usetikzlibrary{patterns}
    
\usepackage{pgfplots}
\usepgfplotslibrary{fillbetween}
\usepackage{adjustbox}
\usepackage{gincltex}

\usepackage{filecontents}
\usepackage{graphicx}
\usepackage{textcomp}
\usepackage{xcolor}
\usepackage{authblk}
\def\BibTeX{{\rm B\kern-.05em{\sc i\kern-.025em b}\kern-.08em
		T\kern-.1667em\lower.7ex\hbox{E}\kern-.125emX}}

\pgfplotsset{compat=1.14}

\usepackage{fancyhdr}
\fancypagestyle{firststyle}
{
	\fancyhf{}
	
	\fancyhead[L]{\scriptsize{This article has been accepted for publication in IEEE Communications Standards Magazine, but has not been fully edited. Content may change prior to final publication.\\
			\copyright 2022 IEEE. Personal use of this material is permitted. Permission from IEEE must be obtained for all other uses, in any current or future media, including
			reprinting/republishing this material for advertising or promotional purposes, creating new collective works, for resale or redistribution to servers or lists, or
			reuse of any copyrighted component of this work in other works.
	}}
}

\begin{document}
\bstctlcite{IEEEexample:BSTcontrol}

\title{Multi-SIM support in 5G Evolution: \\Challenges and Opportunities}

\author{
O. Vikhrova
, S. Pizzi
, A. Terzani
, L. Araujo
, A. Orsino
, G. Araniti
\thanks{
O. Vikhrova, S. Pizzi, and G. Araniti are with the DIIES Dept., University Mediterranea of Reggio Calabria, Italy}
\thanks{
G. Araniti is also with Peoples' Friendship University of Russia (RUDN University), Russia}
\thanks{
A. Terzani and L. Araujo are with Ericsson AB, Sweden}
\thanks{
A. Orsino is with OY L M Ericsson AB, Finland}
\thanks{
A. Orsino is the contact author: Hirsalantie 11, 02420 Kirkkonummi, Finland; e-mail: antonino.orsino@ericsson.com}

}

\maketitle

\thispagestyle{firststyle}

\begin{abstract}
Devices with multiple \gls{sim}s are expected to prevail over the conventional devices with only one SIM.
Despite the growing demand for such devices, only proprietary solutions are available so far. To fill this gap, the \gls{3gpp} is aiming at the development of a unified cross-platform
solutions for multi-SIM device coordination. This paper extends the technical discussion and investigation of the \gls{3gpp} solutions for improving \gls{mt} service delivery to multi-SIM devices. Implementation trade-offs, impact on the \gls{qos}, and possible future directions in \gls{3gpp} are outlined. 
\end{abstract}

\section{Introduction}
\label{sec:introduction}


As smartphones and services became more affordable, their users have tended to use different mobile subscriptions (i.e., SIM cards) for travel, business, and personal needs. For example, the total number of \gls{sim} connections worldwide has exceeded the global population while the number of unique subscriptions make only 60\% of the population~\cite{gsmaeco}. By that time, mobile devices that accommodate more than one \gls{sim} card, also known as dual-\gls{sim} devices, have gained popularity, especially in countries with uneven coverage and in-country roaming. With the advent of \gls{esim}~\cite{gsma}, managing multiple mobile subscriptions within a single device has become more agile and user-friendly. The \gls{esim} technology allows customers to install many \gls{esim} profiles and select between the subscriptions at will via a software menu, although only one profile can be active at any given time. 

The introduction of \gls{esim} facilitates the global adoption of multi-SIM devices~\cite{oasis}. The technology has been adopted by all leading smartphone manufacturers, including Apple, Samsung, and Huawei. 
Since the addition and change of \gls{mno}s became easier and faster, mobile operators expect substantial revenue from providing \gls{esim} and improving dual-SIM user's experience~\cite{simlocal}. 

With multiple subscriptions, \gls{mt} services, such as voice calls, \gls{sms}, and emergency notifications from different networks risk to overlap and fail to reach the user. Therefore, service reception at multi-\gls{sim} devices may require specific software and protocol enhancement on \gls{ue} and network sides. 

So far, device manufacturers have been handling the case independently and in an implementation-specific manner without coordination and support from the \gls{3gpp}. This has led to various proprietary \gls{ue} solutions and network functions implementations with potential network performance degradation and user experience deterioration. 

To fill this gap, \gls{3gpp} has put effort into unifying cross-platform solutions to coordinate multi-\gls{sim} device operation. 
Specifically, it has defined use cases and solution requirements in the study item \cite{3GPP22834}. A consequent study~\cite{3GPP23761} has discussed challenges to support multi-\gls{sim} functions and indicated possible solutions with the aim to ensure reliable and seamless \gls{ue} operation in \gls{4g} and \gls{5g} systems. 
According to the \gls{3gpp} vision, all \gls{sim} registrations of a device should be treated as independent \gls{ue}s from the network perspective. This clearly imposes challenges on the delivery of \gls{mt} traffic over different \gls{sim}s and \gls{ue} mobility management operations.

We aim to analyze the efficacy of solutions for multi-\gls{sim} management 
under investigation at \gls{3gpp} from the radio access and core network perspective. Signaling overhead, implementation cost, and corresponding service latency significantly depend on what solution is adopted in a certain network deployment. Our comparative analysis and considerations about the central \gls{3gpp} solutions contribute to the existing discussion across the industry and standardization bodies. 

First, we introduce the main issues raised by multi-\gls{sim} operation and the motivations behind the \gls{3gpp} study. 
Next, we summarize the state-of-the-art of mobility management and \gls{mt} traffic delivery in \gls{4g} and \gls{5g} systems and overview solutions for the efficient handling of multiple network registrations and service reception from different \gls{mno}s. 
In the end, we discuss the pros and cons of the envisaged system enhancements 
and outline future directions.

\section{Multi-SIM operation issues}
\label{sec:issues}

\begin{figure}
    \centering
    \includegraphics[width=0.9\columnwidth]{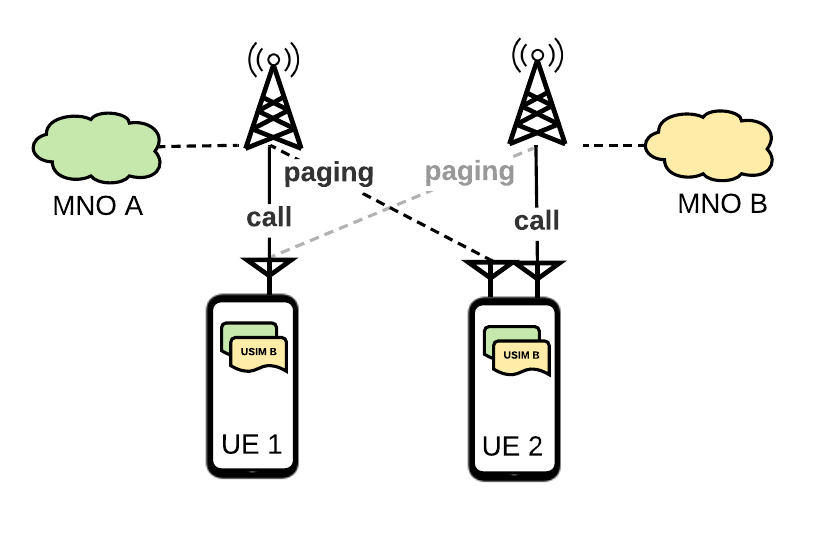}
    \caption{Different configurations of Multi-SIM UEs.}
    \label{fig:multiusim}
    \vspace{-0.5cm}
\end{figure}

Before the \gls{3gpp} study of the multi-\gls{sim} operation related issues in~\cite{3GPP22834}, \gls{gsma} defined a minimum set of requirements intended to ensure consistent behavior of multivendor multi-\gls{sim} devices and gave the definition of a multi-\gls{sim} device~\cite{gsma}. In the context of its technical specification, multi-\gls{sim} devices either have a single \gls{3gpp} network connection and a single \gls{imei} associated with a single selected SIM from several within the device, or have multiple simultaneous \gls{3gpp} network connections and multiple \gls{imei}s, each of which is associated with a particular \gls{sim}. The former means that both \gls{sim}s share the same cellular transceiver, while the latter assumes that each \gls{sim} uses a dedicated transceiver, as illustrated in Fig.~\ref{fig:multiusim}.

The \gls{gsma} has defined \gls{dsds} and \gls{dsda} operation modes for two different categories of multi-\gls{sim} devices. 
\begin{itemize}
    \item \textbf{\gls{dsds}} device (UE 1 in Fig.~\ref{fig:multiusim}) maintains ``idle'' connections to the two networks through time multiplexing. 
    When in-call on one of the \gls{sim}, the connection to the corresponding network moves to the ``active'' state and the \gls{ue} becomes unavailable for paging (unreachable) on another network for the duration of the call.
    \item \textbf{\gls{dsda}} device (UE 2 in Fig.~\ref{fig:multiusim}) can simultaneously maintain ``active'' and ``idle'' connections owning to the dedicated transceiver for each \gls{sim}.
\end{itemize}

To ensure the consistent behavior of different types of multi-\gls{sim} devices in cellular networks, the \gls{3gpp} has defined four issues caused by the lack of unified solutions for multi-\gls{sim} operation.
\begin{itemize}
    \item \textbf{Ongoing service interruption.} While having ongoing service over one \gls{sim}, a \gls{ue} may need to reply to paging message or update its location in the other network. Any transmission associated with one \gls{sim} may cause an ongoing service interruption over the other \gls{sim}. 
    \item \textbf{Unwanted resource wastage.} If a \gls{ue} becomes unavailable for paging in one system due to an active communication in the other one without appropriate network notification, resources allocated for paging will be unused. Moreover, when network fails to reach \gls{ue} in the expected cell, it escalates paging to a wider area, which requires extra resources. 
    \item \textbf{Misleading assumption of reachability.} If a \gls{ue} can not reply to the paging message or the ongoing connection gets disrupted, the system may assume that the \gls{ue}, e.g., has moved to a new \gls{ran} area or experiences low link quality. The wrong assumption leads to statistics distortion misguiding algorithms that rely on them. 
    \item \textbf{Collision of paging occasions.} 
    Paging occasions of a \gls{ue} for different \gls{sim} profiles may overlap either occasionally or systematically. Paging collision can be resolved on different levels of paging management.
    However, any unresolved collisions will cause service disruption or loss, resource wastage, and incorrect assumption of \gls{ue} reachability for paging.
\end{itemize}

Privacy implication issues are currently not in the \gls{3gpp} agenda. However, preliminary ideas on secure communication of multi-\gls{sim}s related information can be found in~\cite{takahito2020ssosiation}.

To deal with the above challenges, the \gls{3gpp} has defined 28 candidate solutions for improved multi-\gls{sim} operation in~\cite{3GPP23761}. These solutions specify methods or necessary changes to the existing procedures to account for the different aspects of the multi-\gls{sim} \gls{ue} configuration, connections onto the networks and user preferences regarding the handling the arriving services when busy with a call over another \gls{sim}. The solutions under \gls{3gpp} discussion have been grouped into different categories summarized in Table~\ref{tab:usecases}. Thus, some solutions are applicable only if connections for all \gls{sim}s are maintained in \gls{5g} systems, while others allow a connection on \gls{4g} networks.  

\begin{table}[htb]
    \centering
    \caption{Classification of multi-SIM support challenges.}
    \begin{tabularx}{\columnwidth}{|p{2cm}|p{6cm}|}
    \hline
        \textbf{Focus} & \textbf{Categories} \\
        \hline
        \gls{cn} connection &  $\bullet$ \gls{ue} connects with \gls{5g} on both \gls{sim}s \\
        & $\bullet$ \gls{ue} connects with \gls{5g} and \gls{eps} \\
        & $\bullet$ \gls{ue} connects with \gls{eps} on both \gls{sim}s \\
         \hline
        \gls{ue} configuration & $\bullet$ single Rx and single Tx
        \\
        & $\bullet$ double Rx and single Tx antenna 
        \\ \hline
        \gls{mno} &  $\bullet$ both \gls{sim}s are owned by the same \gls{mno} \\
        configuration & $\bullet$ \gls{sim}s belong to different \gls{mno}s \\ \hline
        Cell camping & $\bullet$ Multi-SIM \gls{ue} camps on the same network \\
        & $\bullet$ Multi-\gls{sim} \gls{ue} camps on different networks \\ \hline
        Services & $\bullet$ support for \gls{3gpp} services only \\
        & $\bullet$ support for \gls{3gpp} and non-\gls{3gpp} services \\
        & $\bullet$ support for emergency calls \\
    \hline
    \end{tabularx}
        \label{tab:usecases}
\end{table}

\section{Mobile-terminated traffic delivery}
\label{sec:stateoftheart}

Multi-\gls{sim} operation concerns \gls{mm} procedures such as \gls{ue} registration in a \gls{plmn} and its subsequent tracking within the network for the \gls{mt} traffic delivery~\cite{alsaeedy2019tracking}. In \gls{4g}, a \gls{mme} is responsible for the \gls{mm} functions including cell and network selection, \gls{ue} registration, paging, connection establishment and handover. In \gls{5g}, the \gls{mm} is controlled by two separate functions, namely \gls{amf} and \gls{smf}. The former is in charge of managing \gls{ue} authentication, authorization, and mobility control, while the latter mainly provides session management and \gls{ip} address allocation~\cite{gures2020mobility}. 

After the \gls{sim} activation, a \gls{ue} selects a \gls{ran} and camps on a 
\gls{plmn} during the attach procedure. Upon 
registration, \gls{ue} can be either in IDLE or CONNECTED state from the \gls{cn} and \gls{ran} point of view. At the end of the data transmission, registered \gls{ue} is moved to the IDLE state within the \gls{cn} and \gls{ran} 
having logical connection with the \gls{cn} over \gls{nas}. At this stage, the \gls{ue} can only receive signaling messages, 
while its location at the cell level is unknown. 

\begin{figure}
    \centering
    \includegraphics[width=\columnwidth]{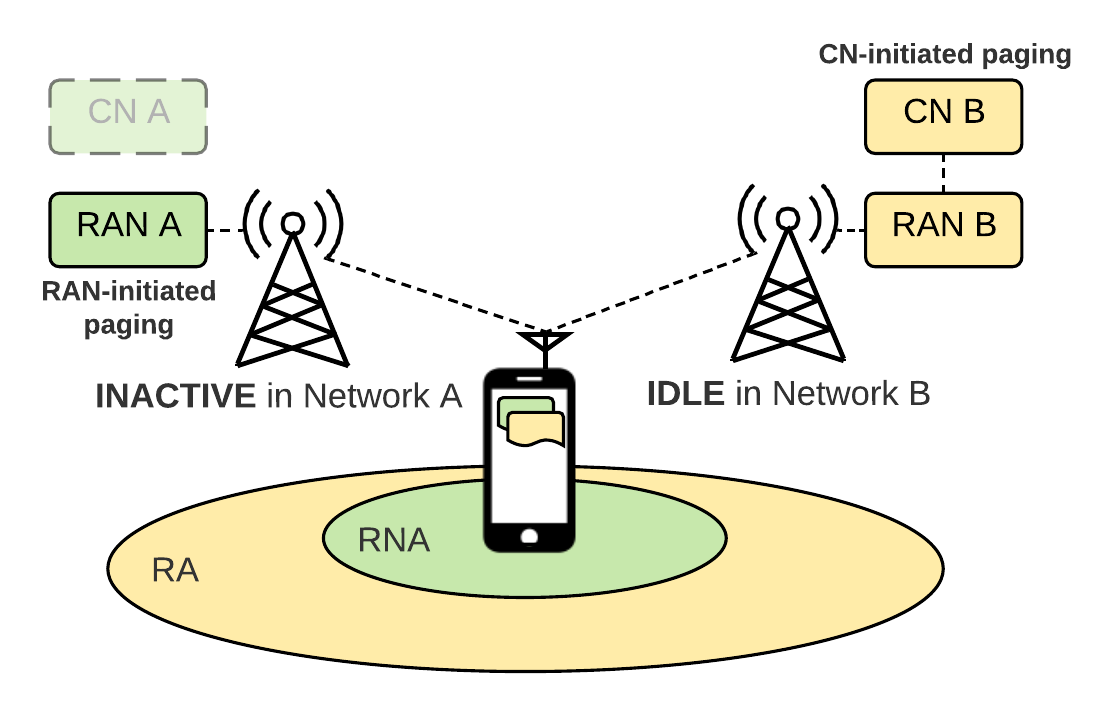}
    \caption{RAN-initiated and CN-initiated paging.}
    \label{fig:paging}
\end{figure}

When a connection between a \gls{ue} and \gls{ran} is established, the \gls{ue} moves to the CONNECTED state in both network segments and 
can be explicitly tracked by the network. In \gls{5g}, besides CONNECTED and IDLE states, a \gls{ue} can be moved to an INACTIVE state, which allows the network to preserve ``active'' connection with \gls{cn}, while the connection within the \gls{ran} is ``idle''. When a \gls{ue} is in the INACTIVE state, it can be searched for in a narrower \gls{ran} area.
Position of an IDLE \gls{ue} in \gls{4g} and \gls{5g} is tracked within the dedicated \gls{ta} and \gls{raa}, respectively. The \gls{ta}/\gls{raa} consists of the list of cell IDs used for the \gls{ue} search. The \gls{raa} includes several \gls{ta}s allowing for steering \gls{ue} search and access in \gls{5g} systems. Every time a \gls{ue} moves outside the \gls{ta}/\gls{raa}, it should update the corresponding \gls{mme}/\gls{amf} about its new location through the \gls{tau} or \gls{rau} (\gls{5g} only) procedure. 
Paging is a prerequisite procedure for any network-originated communication as it helps to determine the \gls{ue} location in the \gls{ran} and subsequently reroute \gls{mt} traffic if the \gls{ue} is in the IDLE state. Unlike the paging procedure in \gls{4g}, which is always triggered by the \gls{cn}, \gls{5g} discerns between CN-originated and RAN-originated paging, as illustrated in Fig.~\ref{fig:paging} and explained in what follows~\cite{hailu2017hybrid}. 

\begin{itemize}
    \item \textbf{CN-originated paging}. To identify the \gls{ue}'s location, \gls{mme}/\gls{amf} cascades a paging message in all cells from the \gls{ue}'s \gls{ta}/\gls{raa} list. Each base station then tries to reach the \gls{ue} at a specific \gls{pf} and \gls{po} inside the frame. The \gls{ue} listens to the paging indication in the downlink at a given \gls{po}.
    When the page is received, the \gls{ue} usually replies, so the \gls{cn} can locate the \gls{ue} on the cell level and forward \gls{mt} traffic to the serving base station. 
    \item \textbf{RAN-originated paging.} If a \gls{ue} is in the INACTIVE state, its location is tracked at the level of \gls{rna}. 
    When \gls{mt} traffic arrives, it is immediately rerouted to the last \gls{ran} used by the \gls{ue}. Then, the \gls{ran} initiates the paging procedure to notify the \gls{ue} about the upcoming downlink transmission.
\end{itemize}

By default, the \gls{pf} and \gls{po} of the \gls{ran}-originated and \gls{cn}-originated paging coincide. To randomize \gls{ue} paging instances in \gls{4g}, the \gls{pf} and \gls{po} values are calculated by an algorithm based on the \gls{imsi}, which is allocated by the \gls{mno}. 
For paging \gls{ue}s in the CONNECTED state, the \gls{mme} allocates a temporally unique ID upon the successful random access attempt. 
To protect \gls{ue}s against the \gls{imsi} catchers and ensure secure paging in 5G, \gls{3gpp} has replaced \gls{imsi} with a secure analogue, known as \gls{suci}. For paging transmission protection, new temporal IDs for \gls{cn}-based and \gls{ran}-based paging have been introduced, which are also used to obtain the \gls{ue} specific \gls{pf} and \gls{po}. It is compulsory to periodically refresh temporal IDs to make it more difficult for an attacker to track a device during paging~\cite{imsicatchers}.

\section{3GPP solutions for Multi-SIM device support}
\label{sec:solutions}

The ability of a \gls{ue} with multiple active \gls{sim}s to properly receive paging notifications from different networks and reply to them in time greatly depends on the \gls{ue} configuration. 
The two most challenging scenarios are when the \gls{ue} can use one Tx and one or multiple Rx antennas. These scenarios are in focus of the \gls{3gpp} study~\cite{3GPP23761}. 

In this section we provide an overview of solutions for the multi-\gls{sim} support in cellular networks discussed in the 3GPP SA2 and RAN2 working groups. 
Solutions outlined as \textit{general methods} are aimed at handling \gls{mt} service, coordinating leaving for another call when only one Tx antenna and any number of Rx antennas are available to a \gls{ue}. These solutions deal with the problem that the \gls{ue} can reply only to one network at a time. Solutions of the second group cover the case when the number of available \textit{Rx antennas} is limited to \textit{one} and \gls{ue}s can not receive paging messages from both networks. The last group is dedicated only to \textit{paging collision avoidance}. These methods can be applied together with the solutions from the other groups to improve the quality of \gls{mt} services. 

\begin{figure*}
    \centering
    \includegraphics[width=\textwidth]{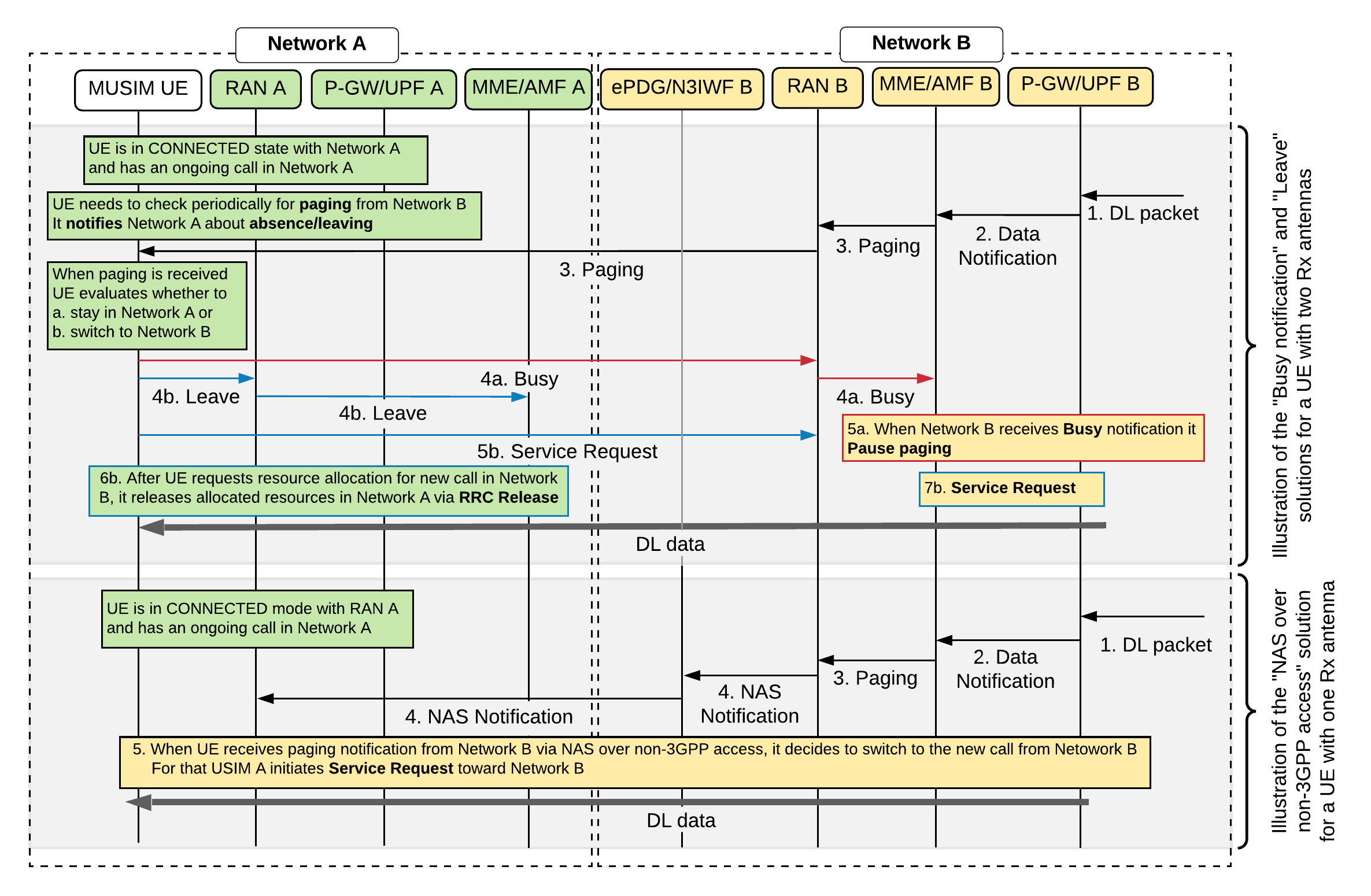}
    \caption{Flow diagram for the Multi-SIM UE coordination over 3GPP accesses and with a non-3GPP access option.}
    \label{fig:flowchart}
    \vspace{-0.5cm}
\end{figure*}

The scenario under consideration is a \gls{ue} with two active \gls{sim}s connected to \gls{4g} and/or \gls{5g} network. The information about the support of the following solutions in these two generations of cellular networks is summarized in Table~\ref{tab:comparison}. A user would typically choose one network as a primary for more important or frequent calls and assume the other network(s) as secondary.
In Fig.~\ref{fig:flowchart} we give an example of the message flow between two networks when a UE has an ongoing call in one network and receives paging from the other one applying several solutions introduced below. 

\subsection{General methods}
\label{sec:solutions:A}

\textbf{Paging Cause.} 
Rules for handling of \gls{mt} service at the multi-\gls{sim} device can be pre-defined by a user or follow some built-in logic. For example, a rule may instruct a device whether it should present a notification about a new voice call to the user if the latter is engaged in another voice call. Similar rules can be defined for the handling \gls{mt} user plane (voice call) and control plane (\gls{sms} or emergency notification) traffic at the network side. A \textit{Paging Cause} can indicate the type of arriving service to assist the \gls{ue} and network with choosing an appropriate strategy for handling the \gls{mt} traffic. It is proposed to include information about the paging cause into the paging message as a new field. In case of the voice call arrival, paging cause should also be included into the Network Triggered Service Request (Downlink Data Notification) initiated by the User Plane Function (Serving Gateway) to inform \gls{amf} (\gls{mme}) about the user plane traffic arrival in \gls{5g} (\gls{4g}). 
The solution requires modifications of the paging message and has \gls{ran} and \gls{cn} impacts since the involved nodes should process the new field. This solution alone lacks feedback when a user receives the page but decides not to respond, which leads to a significant waste of resources, because of the paging escalation. Furthermore, the \gls{amf} (\gls{mme}) may come to an incorrect conclusion about the reason for the paging failure.

\textbf{Notification about short absence period.} 
If a user needs to leave an ongoing call in the primary network for a short time to reply to the page or perform \gls{rau}/\gls{tau} in the secondary one, it can send notification of a short absence to the primary network and communicate with the secondary network without losing allocated resources in the primary one. During the absence, the traffic at the primary network can be delayed, buffered, or discarded according the \gls{qos} requirements. The \gls{ue} does not need to request \gls{rrc} connection release in the primary network for a period of absence. 
This solution aims at avoiding ongoing service interruption and saving network resources and can be applied in \gls{4g} and \gls{5g} networks. 
How to inform the secondary network about the \gls{ue} comeback to the primary network is not specified yet. Furthermore, the solution does not provide any means to leave the primary network when needed. 

\textbf{Busy indication.}
When a \gls{ue} receives a paging message in the secondary network, it can reply with a \textit{busy} notification if the connection with the primary network is preferable. Then the \gls{amf} stops paging and replies to the node initiated paging that the \gls{ue} is reachable, but it has denied the service. This solution excludes paging escalation and prevents the secondary network from misleading assumptions about the \gls{ue} availability. This new feature can be supported in \gls{5gc} and \gls{eps}. However, whether it should be used when \gls{ue} is in the INACTIVE state is not decided yet.

\textbf{Local leaving.} 
When a \gls{ue} decides to leave the primary network, it should properly release allocated radio resources. To minimize signaling overhead and latency, the \gls{ue} can trigger \textit{local leaving} by sending the notification to the \gls{cn} over the \gls{as}. Once the notification is sent, the \gls{ue} can immediately switch to the secondary network, and the \gls{cn} takes responsibility for releasing \gls{rrc} in the primary \gls{ran} and terminating the service. This solution can be implemented both in \gls{5g} and \gls{4g} systems.
To make the solution more efficient, the \gls{ue} should negotiate local leaving configuration with the \gls{ran}. Then the latter can instruct the \gls{ue} whether it should send a release or suspend request (only in \gls{5g}) when leaving the network. 

\textbf{Scheduling gap.}
When only paging detection in the secondary network is required, the \gls{ue} may request a \textit{scheduling gap} to tune away from the primary network for the page reception. The \gls{ran} decides whether the scheduling gap requested through the \gls{as} signaling can be used. The solution can be used when \gls{ue} is in the CONNECTED state in one of the networks. It can not be applied for multi-\gls{sim} \gls{ue} with \gls{4g} only registrations. It does not involve \gls{cn} interaction but could be reasonably used in combination with a busy indication as a reply to the received paging message in the secondary network.

\textbf{Graceful leaving and resumption.}
The solution allows a \gls{ue} to leave the primary network and notify it about the return. As for the local leaving, the \gls{ue} can inform the \gls{cn} over the \gls{nas} including additional information such as expected period of absence and policy for handling \gls{mt} traffic.
The former helps the \gls{ran} to decide whether to move \gls{ue} to the IDLE or INACTIVE state (applicable only in \gls{5g}). If \gls{ue} is in the CONNECTED state, it uses \gls{as} communication to notify the network about leaving and return. 
The solution is aimed at reducing signaling overhead and minimizing the risk of service interruption. It can be enabled in \gls{4g} and \gls{5g} impacting \gls{ran} and \gls{cn} segments. 

\textbf{Leave and return.} 
Another way to avoid uncoordinated leaving from the primary network 
is to indicate over \gls{as} a user's intention to abandon the ongoing service. 
Once confirmed, the \gls{ran} replies back with the \gls{rrc} Release message and moves the \gls{ue} to the IDLE state. The \gls{cn} holds paging for a network-configured time interval and evokes \gls{mt} traffic buffering. At the end of the interval, the \gls{cn} sends the page to the \gls{ue} to re-establish the connection and deliver the buffered traffic. If it does not receive a reply, the paging is considered failed. 
If \gls{ran}-originated paging fails, the buffered data is discarded, and all relevant nodes are informed that the \gls{ue} is unavailable. 



\subsection{Solutions for Multi-\gls{sim} \gls{ue} with single Tx/Rx antenna}
\label{sec:solutions:B}

\textbf{Push notification.}
A \gls{ue} with limited hardware capabilities may indicate to the network A its intention to register for paging events since it can not monitor the paging channel while communicating with the network B. If the \gls{cn} acknowledges the request, it provides the \gls{ip} address of a network function or a node referred to as \textit{Paging Server}. 
When network B wants to page a \gls{ue}, it initiates the direct paging and sends \textit{Push Notification} via Paging Server B. If \gls{ue} is unavailable for the direct paging in network B, it will receive the \textit{Push Notification}. If \gls{ue} is in the IDLE state in both networks, the \gls{ue} will be paged sequentially: first with the direct paging in network B and with the Push Notification in network A after some delay. The arrival of downlink traffic in network A triggers the direct paging in network A. Therefore, the \gls{ue} will be reached in network A if it is not available in network B. These two paging processes are independent. If both successful, the push message will be ignored. This solution is available in \gls{4g} and \gls{5g} networks. 

\textbf{Notification via non-3GPP Access.}
Notification of the \gls{mt} traffic arrival in the secondary network can be delivered in the primary network via the \gls{n3iwf}. The \gls{ue} should indicate in the registration request to be reachable for paging through the \gls{n3iwf}. When \gls{mt} traffic arrives in the secondary network, the \gls{amf} reroutes it to the proper \gls{n3iwf}. The notification from the secondary network will be delivered over the user plane in the primary network to avoid \gls{ue} paging in the secondary network. Thus, the \gls{ue} does not need to switch between networks and deal with any uncoordinated leaving and paging losses. However, the solution requires an initial configuration of the non-\gls{3gpp} access, connection establishment. It is supported only in \gls{5g}.

\textbf{Push notification via SMS.}
The solution is similar to the Push notification and supported by \gls{5gc} and \gls{eps}, but the notification is delivered via \gls{sms}. A \gls{ue} has to indicate to the \gls{cn} of the primary \gls{plmn} that it wants to register for paging events and receive notification via \gls{sms}. This solution does not require any business relationship between the two \gls{plmn}s because the user ID explicitly indicates the identity of the secondary \gls{plmn} to the primary one and vice versa.

\subsection{Solutions for avoiding paging collisions}
\label{sec:solutions:C}

\textbf{NAS parameters change.}
Overlapping \gls{po}s cause paging loss and system performance degradation. For multi-\gls{sim} \gls{ue} in \gls{4g}-only networks, such collision leads to a systematic paging loss, because the \gls{po}s are calculated based on the long-term \gls{imsi}s.  
If \gls{ue} detects \gls{po}s conflict during the registration at the \gls{5g} system, it should request a new \gls{5g-guti} through the mobility update procedure indicating suitable values. If \gls{amf} can not allocate a \gls{5g-guti} that fits the assistance information, it may assign new \gls{po}s by updating parameters for the paging mechanism, e.g., indicating a new paging offset. The natural extension of the previous solution that can be used in any systems is to request an \textbf{alternative UE\_ID} from the \gls{mme} or \gls{amf}. In case of \gls{ran}-based paging, the network accepts the proposed UE\_ID only for calculating a new timing for paging. For the \gls{cn}-based paging, the new UE\_ID can be enforced only after the network confirmation. The new \gls{po} can be used until the \gls{ue} revokes it. Paging collisions could be dynamic because the \gls{ue}-\gls{id} in \gls{5gc} is periodically updated. 

\textbf{Paging offset.}
If two or more \gls{sim} registrations in \gls{eps} result in paging collision, a \textit{multi-\gls{sim} offset} can help to resolve the \gls{po} overlapping when \gls{ue}-\gls{id} can not be changed. The offset value should not lead to even a partial \gls{po} overlapping. Therefore, it should be assigned taking into account the number of active registrations and offset values for all other \gls{sim}s. The solution can also be used in \gls{5g}.

\textbf{Paging on consecutive POs.}
To eliminate the risk of \gls{po} collision when \gls{ue} receives all paging notifications over the primary network, the \gls{po}s of both networks should be consecutive in time. 
The solution proposes to extend the legacy paging with a counter of paging attempts giving a chance to reach the \gls{ue} at the next \gls{po} following the collided one. 


\begin{table}[]
    \centering
    \caption{Overview of the 3GPP solutions for Multi-SIM coordination.}
    \begin{tabular}{|p{4.6cm}|p{0.7cm}|p{0.7cm}|p{0.3cm}|p{0.3cm}|}
    \hline 
    \centering{\textbf{Solutions}} & \textbf{RAN}- based & \textbf{CN}- based & \textbf{5G} & \textbf{4G} \\ \hline 
    1. Paging Cause & \checkmark & \checkmark & \checkmark & \checkmark \\
    2. Short absence  & \checkmark &  & \checkmark & \checkmark \\
    3. Busy  &  \checkmark &  \checkmark & \checkmark & \\
    4. Local leaving  & \checkmark & \checkmark & \checkmark & \checkmark  \\
    5. Graceful leaving  & \checkmark &  \checkmark & \checkmark & \checkmark \\
    6. Leave and return &  & \checkmark & \checkmark & \checkmark \\
    7. Scheduling gap  & \checkmark & \checkmark & \checkmark & \\ \hline
    8. Push notification &  & \checkmark & \checkmark & \\
    9. Notification over non-3GPP access &  & \checkmark & \checkmark  & \checkmark \\
    10. Notification via SMS  & & \checkmark & \checkmark & \checkmark \\ \hline
    11. NAS parameters change  & & \checkmark & \checkmark &  \\
    12. Alternative UE\_ID  &  & \checkmark & \checkmark &  \\
    13. Paging offset & \checkmark & \checkmark & \checkmark & \checkmark \\
    14. Consecutive POs paging & \checkmark & \checkmark & \checkmark & \checkmark \\
         \hline 
    \end{tabular}
    \label{tab:comparison}
\end{table}

\section{Qualitative analysis and open issues}
\label{sec:openissues}

As the above solutions have different impact on \gls{ran}, \gls{cn}, \gls{ue}, 
in this section we analyze their drawbacks, cost of implementation, and discuss possible improvements. We present our qualitative evaluation based on the available 3GPP documents and discussions in SA2 and RAN2 groups, and do not intend to give a definitive conclusion on what solutions shall be applied. With this analysis we aim to give a holistic perspective on the topic as well as to help readers to put discussed solutions into a broader context. 

\subsection{Pros and cons of the solutions for multi-SIM support}

Among the discussed solutions, a new \textit{Paging cause} field helps avoid inquiry to identify why the UE was paged, minimizing any impact on the ongoing service on the other SIM. However, if UE does not respond to the page, the paging request will be escalated, wasting radio resources within the large registration area. The solution enabling an on-off period of absence with the primary network 
prevents unnecessary interruptions of the ongoing service and paging escalation in the secondary network. However, no user control of the solution is described. 
While being away from the primary network for a short period, e.g., for sending busy notification, some service interruption in the primary network may occur. 

Solutions in subsection~\ref{sec:solutions:B} can be thought of as a notification type and fit all types of multi-SIM devices, even the most constrained ones. However, \textit{push notification} requires the implementation of a paging server, while \textit{SMS-based solution} may suffer from uncontrolled and long delays. Importantly, even if a \gls{ue} is paged by two methods in parallel, the solution can not guarantee that either of them will be successful. The interactions between two \gls{plmn}s and traffic forwarding from one \gls{plmn} to another one in several solutions for \gls{ue}s with a single Rx and Tx antenna raise also charging issues. The straightforward solution is to address these issues to the \gls{mno}s and settle it based on the \gls{sla} agreements between the involved \gls{mno}s.

\textit{Paging collisions} could be solved in many ways, as explained in section~\ref{sec:solutions:C}. A group of solutions based on changing \gls{po} every time a collision risk is detected requires extra signaling and, as a consequence, extra resources utilization and latency increase. Changing the paging strategy also helps to deal with overlapping \gls{po}s. However, the gain of such approaches is questionable due to the need for keeping the strategy up-to-date and to the difficulty of designing a paging strategy that can adapt for a general muti-\gls{sim} scenario.

Some solutions, including the one based on the \textit{\gls{ue} assistance information}, are foreseen to increase signaling and device energy consumption since the \gls{ue} is responsible for detecting \gls{po} collisions, providing support information for leaving, and informing network about the current \gls{ue} capabilities and configuration.

Solutions that tackle resource waste and misleading assumption of reachability 
belong to a group of a \textit{coordinated leaving} approaches, which in turn can be divided into: (i) \gls{nas} triggered leaving, (ii) \gls{as} triggered leaving, with and without release assistant information, and (iii) leaving for a given interval. 
\gls{ue} requests to leave or resume a connection via \gls{nas} signaling can be communicated only within the \gls{ran} (if INACTIVE state is supported) or reach the \gls{cn}. Corresponding \textit{graceful leaving} and \textit{local leaving} solutions aim to minimize the overhead needed for coordinated leave and return and minimize service interruption in the primary network. For a short leave, it is more beneficial to move \gls{ue} to the INACTIVE state than to the IDLE state due to the lower latency and signaling cost. 

\gls{as}-level solutions are applied when a \gls{ue} wants to leave ongoing communication in the primary network because the secondary network's upcoming service has higher priority. The fastest way to tune away from the primary network and start receiving service over another is to use local leaving, which does not require network acknowledgment for the \gls{rrc} release. These resource-efficient solutions for coordinated leaving also raise significant concerns about the further actions aligning between UE and network. 


To date (August 2021) SA2 and RAN2 groups have not come to the conclusion on which of the discussed solutions to pursue in addressing the issues defined in the study item~\cite{3GPP23761}. 
Nevertheless, according to the current discussions in 3GPP, the industry have given preference to the ``local leaving", ``paging cause", and ``paging offset" solutions. 
However, the official updated version of the 3GPP study item will be available only in Q1 2022.

\subsection{CN-based vs RAN-based solutions and steps ahead}
\label{sec:futuredirections}

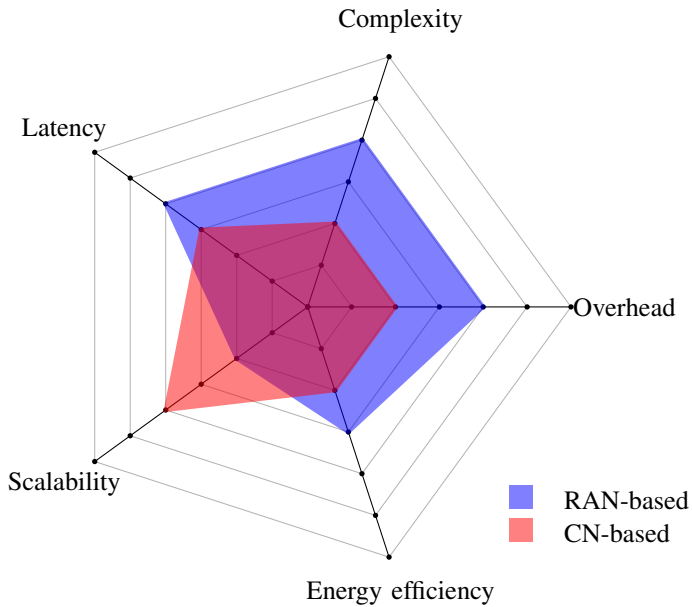
\begin{figure}
    \centering
    \definecolor{green}{rgb}{0.62, 0.84, 0.43}%

\newcommand{\D}{5} 
\newcommand{\U}{6} 
\newcommand\ColorBox[1]{\textcolor{#1}{\rule{2ex}{2ex}}}

\newdimen\R 
\R=3.5cm 
\newdimen\L 
\L=4cm

\newcommand{\A}{360/\D} 

\begin{tikzpicture}[scale=1]
  \path (0:0cm) coordinate (O); 

  \foreach \X in {1,...,\D}{
    \draw (\X*\A:0) -- (\X*\A:\R);
  }

  \foreach \Y in {0,...,\U}{
    \foreach \X in {1,...,\D}{
      \path (\X*\A:\Y*\R/\U) coordinate (D\X-\Y);
      \fill (D\X-\Y) circle (1pt);
    }
    \draw [opacity=0.3] (0:\Y*\R/\U) \foreach \X in {1,...,\D}{
        -- (\X*\A:\Y*\R/\U)
    } -- cycle;
  }
%
  \path (1*\A:\L) node (L1) {Complexity};
  \path (2*\A:\L) node (L2) {Latency};
  \path (3*\A:\L) node (L3) {Scalability};
  \path (4*\A:\L) node (L4) {Energy efficiency};
  \path (5*\A:\L+6) node (L5) {Overhead};

  %
  %

  %
  \draw [color=blue,line width=1.5pt,opacity=0.5,fill=blue]
    (D1-4) --
    (D2-4) --
    (D3-2) --
    (D4-3) -- 
    (D5-4) -- cycle;

  %
  \draw [color=red,line width=1.5pt,opacity=0.5,fill=red]
    (D1-2) --
    (D2-3) --
    (D3-4) --
    (D4-2) --
    (D5-2) -- cycle;
   
\node[anchor=south west,xshift=-70pt,yshift=20pt] at (current bounding box.south east) 
{
\begin{tabular}{@{}lp{3cm}@{}}
\ColorBox{blue!50} & RAN-based \\
\ColorBox{red!50} & CN-based \\
\end{tabular}
};
    \end{tikzpicture}
    \caption{Comparison of the solutions for multi-\gls{sim} coordination.}
    \label{fig:comparison}
\end{figure}

The central discussion of the candidate solutions for the multi-SIM support in Release 17 comes to the question whether the overhead related to handling paging collision and connectivity interruption should be on the \gls{cn} or \gls{ran} side. In this section, we provide a 
concise technical analysis of the RAN-based and CN-based solutions together with the 
future research directions. 
Due to the complexity of the system to be modeled and lack of information on how the different solutions are implemented in reality, a system level simulation was not an appropriate methodology. For this reason, we have modeled each solution according to our assumptions on, e.g., what messages are exchanged between network and multi-SIM UE, what signaling latency is involved. We have drawn conclusions in a form of key performance metrics defined below for the RAN-based and CN-based solutions and summarized them in Figure~\ref{fig:comparison}:
\begin{itemize}
    \item \textbf{Complexity}: to demonstrate at what extent a solution impacts UE, RAN, or CN;
    \item \textbf{Overhead}: to show how much signaling overhead the solution brings;
    \item \textbf{Scalability}: to indicate whether the solution is applicable for a large number of UEs and/or RAN nodes;
    \item \textbf{Latency}: to compare how long it takes for the solution to take effect; 
    \item \textbf{Energy efficiency}: to demonstrate whether a solution can increase energy consumption.
\end{itemize}

The advantage of  \gls{ran}-deployed solutions is the low latency since the signaling happens between two different RANs 
or between UE and 
RAN. 
However, this benefit comes with the assumption that the UE is capable of keeping active RRC connections with both RANs 
simultaneously. 
The 
drawback of 
the RAN-based solution is 
the requirement to perform \gls{mm} procedures in two RANs at the same time. 
It will be particularly challenging for these RANs to 
avoid \gls{po} collisions 
because of the 
network dis-synchronization and the frame structure misalignment (i.e., when the time instants of the frames start and the frames end are different). 

CN-based solutions instead have no impact on the RAN as all the management is done in the CN. A UE can be connected to only one RAN at a time. The complexity and implementation cost of the CN-based solutions are lowest. Furthermore, the UE will be able to save battery since it has to listen to the paging only in one RAN. 
However, the main drawback of the CN-based solutions is an increased latency for the overall procedure with respect to a RAN-based solution.

Currently, 3GPP is analyzing both 
solution categories to identify the best one for providing necessary functions for the multi-\gls{sim} support that introduce lower overhead and system impact. 
The overall picture and 
3GPP 
standardization dynamics lead to the conclusion that CN-based solutions are seemed to be the most suitable as a way forward. Nevertheless, 
since the 
evaluation of all the solutions is still ongoing in 3GPP, a clear understanding of what 
approaches will be adopted 
becomes available only in the first quarter of 2022.

\section{Conclusions}
\label{sec:conclusions}

In this paper, we have highlighted the challenges raised by the global adoption of devices with multiple SIMs and presented approaches for the multi-SIM device coordination investigated by the 3GPP. In particular, we have analyzed the candidate solutions for Rel-17 classified as (i) general methods that build the basis for multi-SIM UE support, (ii) solutions for constrained \gls{ue}s with latency and reliability limitations, and (iii) different solutions for avoiding paging collisions. We have discussed their benefits, drawbacks and impact to support the ongoing work in the 3GPP towards the adoption of CN-based solutions.



\section*{Acknowledgment}
This work has been supported by the RUDN University Strategic Academic Leadership Program.

\renewenvironment{IEEEbiography}[1]
{\IEEEbiographynophoto{#1}}
{\endIEEEbiographynophoto}

\begin{IEEEbiography}
{Olga Vikhrova} (olga.vikhrova@unirc.it) is a PhD Candidate in Information Engineering at the University Mediterranea of Reggio Calabria. She received her B.Sc and M.Sc degrees in Information and Computer Science from the Peoples' Friendship University of Russia (RUDN University) in 2012 and 2014, respectively. Her current research interests include distributed edge learning and computing, convergence of terrestrial and non-terrestrial networks.
\\
\\
\textbf{Sara Pizzi} (sara.pizzi@unirc.it) is an assistant professor in telecommunications at University Mediterranea of Reggio Calabria, Italy, where she received the Ph.D. degree (2009) in Computer, Biomedical and Telecommunication Engineering. Her current research interests focus RRM for multicast service delivery, D2D and MTC over 5G networks, integration of NTN in IoT.
\\
\\
\textbf{Alessio Terzani} (alessio.terzani@ericsson.com) is a System Developer at Ericsson in Sweden. He is currently working in 5G traffic control area focusing on the RAN product development and on the standardization of the RAN node interfaces and protocols, in particular the radio resource control protocol. He also worked in 3G network, in particular in the mobility area and end-to-end testing. He holds a Master's Degree in electronics engineering from University “Sapienza”, Rome, Italy
\\
\\
\textbf{Lian Araujo} (lian.araujo@ericsson.com) is a Senior Researcher at Ericsson in Sweden. He is currently working as a 3GPP delegate in the specification of radio interface protocols.
His areas of interest are mostly related to 5G and the radio resource control protocol, in topics such as carrier aggregation and dual connectivity.
Lian Araujo also performed research in information centric networking and quality of service for mobile networks, previously to Ericsson. He holds an M.Sc. in electrical engineering from Federal University of Pará (UFPA), Belém, Brazil.
\\
\\
\textbf{Antonino Orsino} (antonino.orsino@ericsson.com) is currently a Senior Research at Ericsson Research, Finland, and an Ericsson 3GPP delegate in the RAN2 WG. He holds a Ph.D. from University Mediterranea of Reggio Calabria, Italy. He is actively working in 5G NR standardization activities and he is the inventor/co-inventor of 100+ patent families/applications, as well as the author/co-author of 60+ international scientific publications and standardization contributions in the field of wireless networks. He received the Best Junior Carassa Award in 2016 as the best Italian junior researcher in Telecommunications. 
\\
\\
\textbf{Giuseppe Araniti} (araniti@unirc.it) received the Ph.D. degree in electronic engineering in 2004 from the University Mediterranea of Reggio Calabria, Italy, where he is Assistant Professor of telecommunications. His major area of research is on 5G/6G networks and includes personal communications, enhanced wireless and satellite systems, traffic and radio resource management, eMBMS, D2D and M2M/MTC.
\\
\\
\end{IEEEbiography}

\bibliographystyle{IEEEtran}
\bibliography{multisim_paging.bib}

\end{document}